\documentclass[aps,prl,twocolumn,showpacs,superscriptaddress,preprintnumbers,amsmath,amssymb]{revtex4}

\usepackage{graphicx}
\usepackage{dcolumn}
\usepackage{bm}
\usepackage{amsmath}
\usepackage{siunitx}
\usepackage{graphicx}
\usepackage{subfigure}
\usepackage{natbib}
\usepackage{float}
\usepackage{xcolor}

\begin{document}

\title{Relativistic Oscillating Window Driven by an Intense Laguerre Gaussian Laser Pulse}

\author{Yao Meng}
\affiliation{State Key Laboratory of Dark Matter Physics, Key Laboratory for Laser Plasma (Ministry of Education), Tsung-Dao Lee Institute $\&$ School of Physics and Astronomy, Shanghai Jiao Tong University, Shanghai 201210, China}
\affiliation{Collaborative Innovation Center of IFSA (CICIFSA), Shanghai Jiao Tong University, Shanghai 200240, China}
\author{Runze Li}
\affiliation{State Key Laboratory of Dark Matter Physics, Key Laboratory for Laser Plasma (Ministry of Education), Tsung-Dao Lee Institute $\&$ School of Physics and Astronomy, Shanghai Jiao Tong University, Shanghai 201210, China}
\author{Longqing Yi}
\thanks{lqyi@sjtu.edu.cn}
\affiliation{State Key Laboratory of Dark Matter Physics, Key Laboratory for Laser Plasma (Ministry of Education), Tsung-Dao Lee Institute $\&$ School of Physics and Astronomy, Shanghai Jiao Tong University, Shanghai 201210, China}
\affiliation{Collaborative Innovation Center of IFSA (CICIFSA), Shanghai Jiao Tong University, Shanghai 200240, China}

\date{\today}

\begin{abstract}
	High-order harmonic generation by the diffraction of an intense Laguerre-Gaussian (LG) laser beam through a small aperture is studied.
	It is found that the 2D peripheral electron dynamics on the rim can facilitate complex interplay between the spin and orbital angular momentum interaction, which leads to distinct selection rules for LG pulses with different polarization states.
	In particular, when the driver is linearly polarized, the harmonic beams no longer follow a simple orbital angular momentum conservation rule. Instead, multiple LG modes with different topological charges are produced in each harmonic beam, and the number of modes equals to the harmonic order.
	A theory is derived and validated by simulations, which can predict the harmonic topological charges as well as their relative intensities for LG drivers with different polarization states.
	Our work provides fundamental insight into the behavior of light in nonlinear optics, and paves the way towards high-intensity UV or X-ray pulses carrying controllable OAM, that can serve as versatile tools at frontiers of various scientific fields.
\end{abstract}

\maketitle

Advances in modern laser technology have resulted in exciting opportunities of studying chirality and symmetry with finely controlled laser pulses \cite{Habibovic2024}. In particular, immense effort has been dedicated to manipulate the angular momentum of a laser beam, which can take two forms: spin angular momentum (SAM), associated with the right- or left-handedness of circular polarization, and orbital angular momentum (OAM), corresponding to the vorticity or global helical phase front of light \cite{Bliokh2015a,Shen2019}.
These advanced light sources allows one to gain control of symmetry (and asymmetry) properties of light, which are especially interesting for the research of chiral phenomena in light-matter interaction, such as circular dichroism \cite{Berova2013}, helical dichroism \cite{Brullot2016,Forbes2018,Rouxel2022}, and light-induced chiral manipulation \cite{Mayer2022,Ordonez2019,Wanie2024}.

Among these fields is high-order harmonic generation (HHG) with laser beams carrying OAM \cite{Gariepy2014,Geneaux2016}, which has attracted extensive attention in recent years. 
The selection rules of harmonic topological charge in the HHG process not only reveal the topological nature in nonlinear light-matter interactions \cite{Alon1998,Neufeld2019,Tzur2021}, but also promise a pathway to imprinting OAM to EUV or X-ray light.
Typically, HHG is triggered by laser-gas interaction at low-intensity regime \cite{Corkum2007}, and by so-called ``relativistic oscillating mirror" (ROM) mechanism \cite{Bulanov1994,Lichters1996,Baeva2006,Dromey2006} at high intensities via laser-solid interaction. In both cases, it is found the harmonic optical vortices follow simple OAM conservation rules for the harmonic frequency up-conversion process \cite{HG2013,Zhang2015}.

Recently, a new type of HHG mechanism is proposed based on the diffraction of a relativistically-strong laser pulse, namely the ``relativistic oscillating window" (ROW) \cite{Yi2021,Yi2025,Hu2025}, where the electrons on the rim of aperture quiver in the incident laser field, resulting in relativistic Doppler effect and HHG in the diffracted fields. 
Importantly, comparing to the ROM mechanism, 
the electron dynamics 
is intrinsically 2D within the diffraction screen, and the pattern of electron motion can be imprinted on the optical properties of the harmonic beams.
Therefore the ROW mechanism exhibits rich interplay between the SAM and OAM or light, leading to complex selection rules that essentially reflects the symmetry of the system.
For instance, pinhole diffraction of circularly polarized (CP) light produces harmonic optical vortices \cite{Yi2021}; and by engineered diffraction apertures with n-fold rotational symmetry, a controlled frequency comb with specific OAM can be produced \cite{Trines2024}.
Despite these rapid progresses, to our knowledge, the studies of relativistic diffraction have been restricted to standard planar wave lasers, where the electron dynamics is relatively simple.
To exploit the full degree of freedoms enabled by the 2D peripheral electron dynamics, structured lights, such as Laguerre-Gaussian (LG) beams \cite{Allen1992,Shi2014}, vector beams \cite{RG2018} and spatiotemporal optical vortices \cite{Hu2025,Bliokh2015PR}, should be employed to facilitates control and understanding of the HHG process in relativistic diffractioin.

\begin{figure*}[!t]
	\centering
	\includegraphics[width=0.9\textwidth]{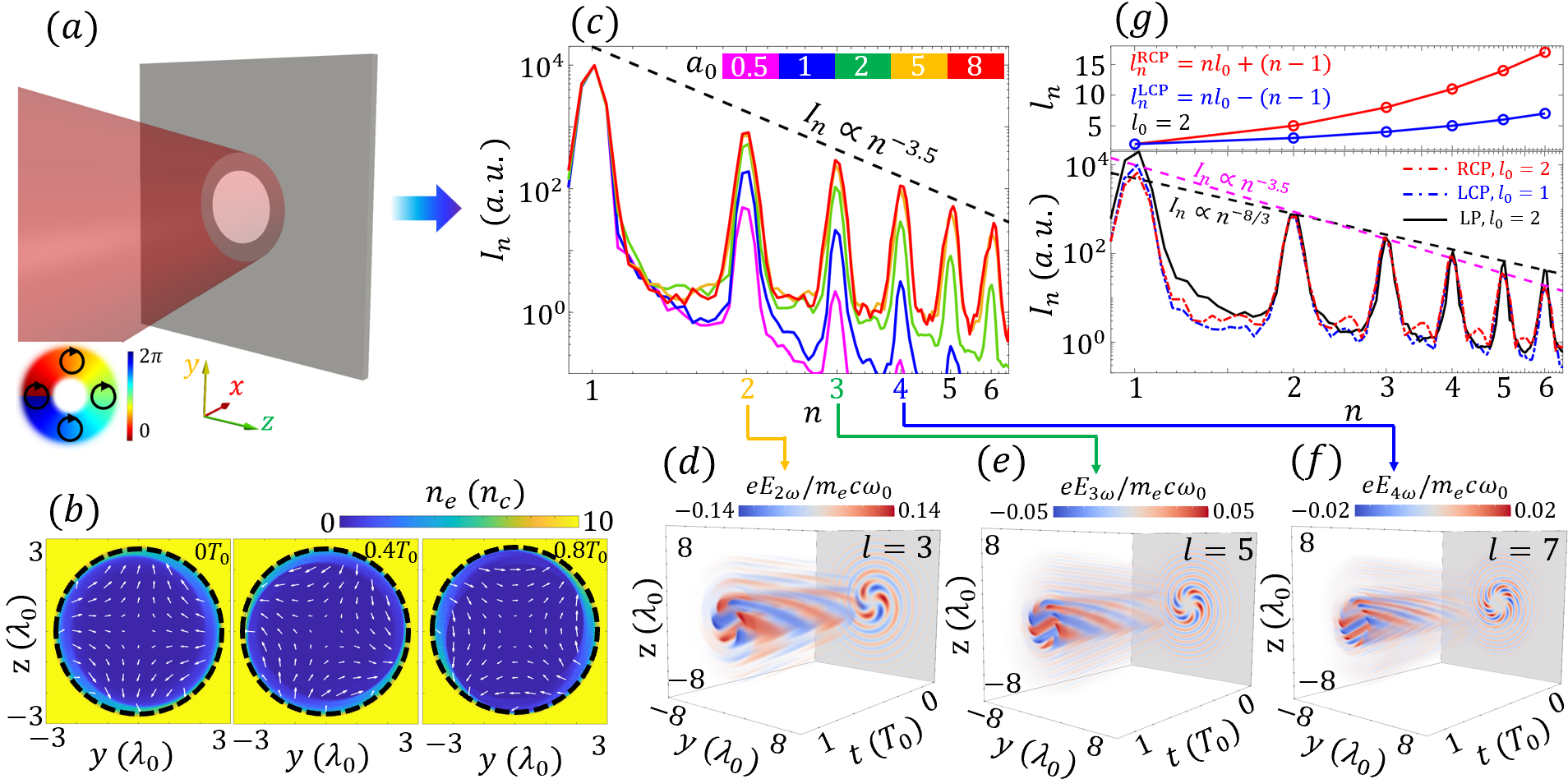}
	\caption{(a) The schematic sketch of a RCP LG pulse with $l_0 = 1$ diffracting through a small aperture on a plasma target. (b) Dynamic electron density distribution on the rim, the three snapshots are separated by 0.4 $T_0$, from left to right, the white arrows represent laser electric field, and the black dashed line indicates the initial boundary. (c) The harmonic spectra produced by a RCP pulse with $l_0 = 1$, the colors represent different drive laser intensity, and the black dashed line shows the fitting by $I_n\propto n^{-3.5}$. The color-coded 3D field distribution of the (d) second-, (e) third-, and (f) fourth-order harmonic beams (with a duration $\sim 1T_0$ is presented) showing they are LG modes with topological charges of $3$, $5$, and $7$, respectively. (g) The harmonic topological charges (upper panel) and the HHG spectra (lower panel) produced by LG pulses with different polarization and topological charges (specified in the plot). We use $l_0 = 2$ in the upper panel, $a_0 = 5$ for CP and $a_0 = 5\sqrt{2}$ for LP drivers in the lower panel of (g), the magenta and black dashed lines show the fitting of HHG spectra by $I_n\propto n^{-3.5}$ and $n^{-8/3}$, respectively.}
	\label{fig:1}
\end{figure*}

In this letter, we investigate the selection rules in the ROW process driven by intense LG laser pulses with different polarization states.
By means of 3D particle-in-cell (PIC) simulations, we found that the topological charges of the harmonic beams depend crucially on the symmetry. 
In particular, when a CP LG pulse is adopted, the system has continuous rotational symmetry, the diffracted lights at a given harmonic order ($n$) consists of a single LG mode with well-defined topological charge ($l_n$). 
By breaking such symmetry with a linearly polarized (LP) drive laser, a family of LG modes are generated at each harmonic order.
The observed selection rules are interpreted by a theoretical model that we present for the first time.\\

\begin{figure*}[!t]
	\centering
	\includegraphics[width=0.9\textwidth]{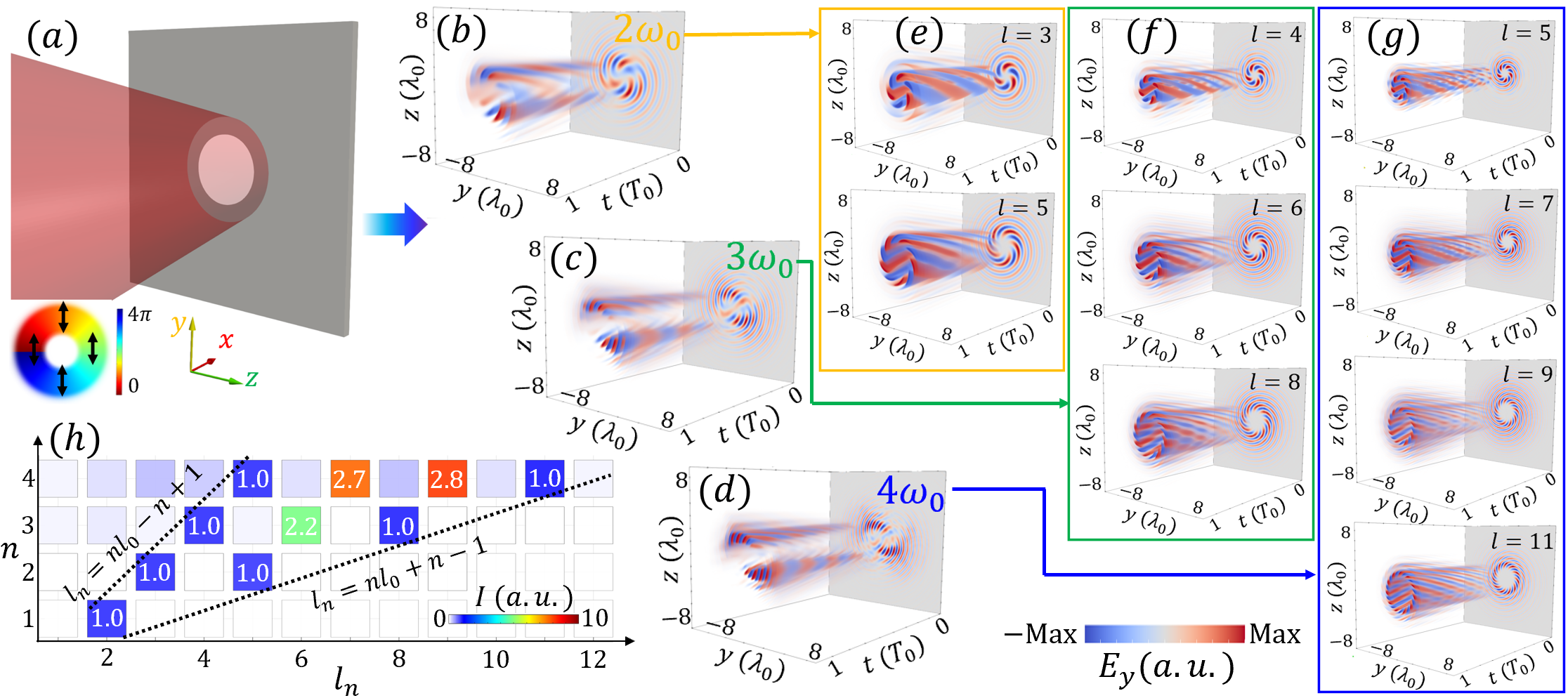}
	\caption{(a) The schematic sketch of a linearly polarized LG pulse with $l_0 = 2$ diffracting through a small aperture. The (b) second- (c) third- and (d) fourth-harmonic beams can be decomposed into a series of LG mode components, as illustrated in (e-g), corresponding to the second-, third-, and fourth-harmonics respectively. (h) The azimuthal spectra of the HHG modes. At each harmonic order $n$, the intensity of the modes are normalized by the lowest-order mode $l_n = nl_0 - n + 1$ produced at $n\omega_0$. The numbers in white show the square root of the relative intensities.}
	\label{fig:2}
\end{figure*}

We first present our simulation results on the diffraction of a CP LG laser. The schematic map of the setup is illustrated in Fig.~1(a), where the drive laser pulse shines through a pinhole on a plasma foil and propagates to $+x$.
The drive laser pulse is expressed by $\mathbf{E_l}=(\mathbf{e_y}+i\sigma\mathbf{e_z})E_0(\sqrt{2}r/w_0)^{l_0}\exp(-r^2/w_0^2)\sin^2(\pi t/\tau_0)\exp(ik_0x-i\omega_0t+il_0\phi)$, for $0<t<\tau_0 = 54$ fs, where $\mathbf{e_y}$ ($\mathbf{e_z}$) denotes the unit vectors in $y$($z$) direction, $r = \sqrt{y^2+z^2}$ is the radial coordinate, $\phi$ is the azimuthal angle measured from $y$ axis.
The normalized laser amplitude is $a_0\equiv eE_0/m_ec\omega_0=2$, with $e$, $m_e$, $c$, $E_0$, and $\omega_0$ being the elementary charge, electron mass, vacuum light speed, laser amplitude and frequency, respectively.
The laser spot size is $w_0 = 4~\si{\micro\meter}$, wave number $k_0=2\pi/\lambda_0$, and wavelength $\lambda_0=1~\si{\micro\metre}$ ($T_0\approx3.3$ fs is the laser period).
The OAM and SAM of the drive laser pulse are controlled by topological charge $l_0$ and $\sigma$, where $\sigma=0$, $+1$, and $-1$ corresponding to LP, right-handed CP (RCP), and left-handed CP (LCP) pulse, respectively.
Thus, each fundamental laser photon carries a total angular momentum of $(l_0+\sigma)\hbar$.
The diffraction screen [assumed aluminum] is modeled by pre-ionized plasma placed at $x_0 = 4~\si{\micro\metre}$, with thickness $l_t = 1~\si{\micro\metre}$ and electron density $n_0 = 30 n_c$, where $n_c=m_e\omega_0^2/4\pi e^2\approx 1.1\times10^{21}~\si{\centi\metre}^{-3}$ is the critical density. A pinhole (radius $r_0 = 3~\si{\micro\metre}$) is present at the center of the screen, and the density profile at the inner boundary ($r<r_0$) is $n(r) = \exp[(r_0-r)/h]$, with $h = 0.2~\si{\micro\metre}$ the scale length. 
The simulations are performed with 3D fully kinetic PIC code {\sc EPOCH} \cite{Arber2015}, The simulation domain is $L_x \times L_y \times L_z = 15~\si{\micro\meter}\times 12~\si{\micro\meter}\times 12 \si{\micro\meter}$, which are sampled by
$1500 \times 600 \times 600$ cells with $10$ macroparticles for
electrons, and $3$ for Al$^{3+}$ per cell. 

When the drive pulse travels through the aperture, the electrons at boundary are wiggled by the strong laser field, resulting in a dynamic diffraction window as shown in Fig.~1(b). The drive laser pulse employed here is a RCP LG beam with $l_0 = 1$.
One can see that the shape of the aperture is deforming as it oscillates, indicating the electron dynamics is controlled by the azimuthal dependence of the laser field [white arrows in Fig.~1(b)].

The electromagnetic waves propagating near the rim of the diffraction window thus interact with this oscillating layer of electrons, leading to HHG. The spectra are presented in Fig.~1(c), which are obtained from PIC simulations by Fourier transformation of the diffracted field recorded at $10~\si{\micro\metre}$ behind the target. One can see that the HHG spectral profiles hardens as the drive laser intensity increases, this trend becomes saturated at $a_c\approx 5$, and the harmonic spectra for $a_0\geq 5$ are almost the same, which can be fitted by $I_n\propto n^{-3.5}$.

The characteristics of the harmonic beams are associated with the electron dynamics, which are controlled by the polarization and phase front of the drive laser pulse. Consequently, the the SAM and OAM of the driver can be imprinted on the harmonic beams.
This is illustrated in Fig.~1(d-f), where the color-coded second-, third-, and forth-order harmonics are displayed in 3D space $(t,y,z)$.
Here each harmonic with order $n$ is obtained by spectral filtering in the frequency range $[n-0.5 ,n+0.5] \omega_0$ from the recorded diffracted field ($E_y$). 
All the harmonic beams show helical phase fronts in connection with $n$, indicating spin-orbital angular momentum interaction takes place.

The harmonic spectra obtained from high-intensity ($a_0>a_c$) LG drivers with different $l_0$ and $\sigma$ are compared in Fig.~1(g). In the lower panel, the HHG spectra of three cases are presented, a RCP driver ($\sigma = 1, l_0 = 2$), a LCP driver ($\sigma = -1, l_0 = 1$), and a LP driver ($\sigma = 0, l_0 = 2$).
Apparently, they show similar power-law shapes, and the topological charge $l_0$ has little influence on the spectral index. 
Meanwhile, the HHG spectrum produced by the LP laser is slightly harder ($I_n\propto n^{-8/3}$) than the CP drivers. This can be attributed to the high-frequency ($2\omega_0$) oscillating terms in the LP laser ponderomotive force, in consistence with our previous study \cite{Yi2025}.

The upper panel of Fig.~1(g) shows the selection rule of the harmonic topological charge ($l_n$) for CP drivers, where we use $l_0=2$ as an example in both RCP and LCP cases. One can see that only one LG mode is generated at each harmonic order $n$, and the topological charge is $l_n = nl_0+(n-1)\sigma$, indicating total angular momentum conservation.\\

When the drive LG pulse is linearly polarized, it breaks the continuous rotational symmetry, which potentially leads to different HHG selection rule. 
We thus perform 3D PIC simulation with the setup in Fig.~2(a). The parameters are similar to that in Fig.~1, but here we employ a LP ($\sigma = 0$) driver with $a_0 = 2\sqrt{2}$ and topological charge $l_0 = 2$. Our key results are summarized in Fig.~2.

To analysis the selection rules for the topological charges in this case, we first conduct Fourier transformation to obtain the 3D harmonic fields. As shown in Fig.~1(b-d), all the harmonics exhibit complex structure, indicating superposition of multiple LG modes.
We then proceed by azimuthal Fourier transforming of each harmonic to yield the topological charges ($l_n$) spectra. 
The single-mode harmonic vortices are subsequently reconstructed via azimuthal filtering of a single $l_n$, as presented in Fig.~2(e-g) for the second-, third-, and fourth-harmonics, respectively.
Apparently, th $n$th-harmonic field consists of $n$ LG modes with different $l_n$, which are centered around $nl_0$, and spaced by $2$.

The relative intensities of the harmonic LG modes are illustrated in Fig.~2(h). One can see that they are distributed within a half-infinite triangle in the $n - l_n$ space. The upper and lower boundaries of the triangle are $l_n = nl_0\pm(n-1)$, corresponding to the selection rules for RCP and LCP cases, respectively. 
Note that at each harmonic order $n$, the intensities of the LG modes are normalized by the lowest azimuthal mode produced with $l_n = nl_0-(n-1)$.
In general, the intensity of the modes are higher in the center, and almost symmetrically distributed with respect to the middle line $l_n = nl_0$. 
Finally, it is worth noting that the square root of the relative intensities (corresponding to the mode amplitudes) are very close to the ``Pascal's triangle", as specified by the white numbers Fig.~2(h).\\

In the following we derive the selection rules based on the framework of ROW model. We first use the Kirchhoff integral theorem and take into account the retarded effects due to oscillating window, and write the diffracted electric fields as \cite{Yi2021,Yi2025}.
\begin{equation}
	\begin{aligned}
	\mathbf{E_{hhg}}(x,y,&z,t)=\frac{1}{2\pi}\nabla \times \int_{B}^{} \lbrack \mathbf{e_{n}}\times\mathbf{E_l}(x_0,y_s,z_s)\rbrack\\
	&\times\frac{{\rm exp}(ik_0R'-i\omega_0t)}{R'} \, ds',
	\end{aligned}
\end{equation}
where $\mathbf{e_{n}}$ is the unit vector normal to the screen, and $R'(t')=|\mathbf{R}-d\mathbf{R'}(t')|$ is the distance between an area element $[ds'(x_0, y_s, z_s)]$ to an observation point $(x, y, z)$, with $\mathbf{R}$ being the initial distance. Here, $d\mathbf{R'}(t') = -(\mathbf{e_y}+i\sigma\mathbf{e_z})\delta_{ow}\exp(ik_0x_0-i\omega_0t'+il_0\phi_s)$ is the displacement of $ds'$ measured at retarded time $t'=t-R'/c$, and $\phi_s = \tan^{-1}(y_s/z_s)$ is the azimuthal angle at the diffraction target. The amplitude of electron oscillation is $\delta_{ow}$, which is assumed to be small ($\delta_{ow}\ll\omega_0/c$) in our derivation of the selection rules, valid when $a_0\ll1$.
Importantly, since only the electromagnetic wave propagating near the oscillating rim of the aperture is relevant for HHG, the integration is over a narrow bound $(B)$ at the boundary ($\sim \delta_{ow}$), thus $ds'\approx\delta_{ow}r_0d\phi_s$. 
Such a quasi-1D area can be twisted to account for the deformation of diffraction window induced by a LG driver [Fig.~1(b)]. 

In order to explain the selection rules, it is sufficient to derive analytically the lowest order of diffracted fields, namely, we take $t'\approx t-R/c$ and substitute it into the displacement of $ds'$ to obtain
\begin{equation}
	\begin{aligned}
		R'(t')&\approx R_0-r_0\sin\theta_0\cos(\phi-\phi_s)+\sin\theta_0\delta_{ow}\\
		\times&(\cos\phi+i\sigma\sin\phi)\exp(ik_0R-i\omega_0t+il_0\phi_s),
	\end{aligned}
\end{equation}
where $\theta_0 = \sin^{-1}(\sqrt{y^2+z^2}/R_0)$ and $\phi = \tan^{-1}(y/z)$ are the opening angle (measured at the center of the diffraction aperture) and the azimuthal angle of the observe point, with $R_0 = \sqrt{(x-x_0)^2+y^2+z^2}$. Since $\delta_{ow}\ll r_0\ll R_0$, all the terms that proportional to $\delta_{ow}r_0/R_0^2$ and smaller are neglected.

For a CP driver, by substituting Eq.~(2) and $\mathbf{E_l}\propto\exp(il_0\phi_s)$ into to Eq.~(1), and applying the Jacobi-Anger identity twice, the diffracted harmonic fields for a CP driver can be obtained as:
\begin{equation}
	\begin{aligned}
		\mathbf{E_{hhg}^{CP}}&\propto\sum_n\sum_mJ_{n-1}(\epsilon_1)J_m(n\epsilon_2)\\
		&\times\exp\{ink_0R_0-in\omega_0t+i[m+(n-1)\sigma\phi]\}\\
		&\times\int_{0}^{2\pi}\exp{[i(nl_0-m)\phi_s]}d\phi_s,
	\end{aligned}
\end{equation}
where $J_n(x)$ is the Bessel function of the first kind, $\epsilon_1=k_0\sin\theta_0\delta_{ow}$ and $\epsilon_2=k_0\sin\theta_0r_0$. Apparently, the integral in Eq.~(3) is nonzero only when $m=nl_0$. Therefore, only one mode is produced at each harmonic with topological charge is $l_n = nl_0+(n-1)\sigma$, in consistent with Fig.~1.

Similarly, substitute $\sigma = 0$ into Eq.~(2), one can obtain $\mathbf{E_{hhg}}\propto\sum_nJ_{n-1}(\epsilon_1\cos\phi)J_{nl_0}(n\epsilon_2)\exp(ink_0R_0-in\omega_0t+inl_0\phi)$ for LP drivers. 
Since $\epsilon_1\ll1$ and $\cos\phi=[\exp(i\phi)+\exp(-i\phi)]/2$, one can write $J_{n-1}(\epsilon_1\cos\phi)\approx\epsilon_1^{n-1}[\exp(i\phi)+\exp(-i\phi)]^{n-1}/[4^{n-1}(n-1)!]$, which yields:
\begin{equation}
	\begin{aligned}
	\mathbf{E_{hhg}^{LP}}&\propto\sum_n\sum_{q=0}^{n-1}\frac{\epsilon_1^{n-1}J_{nl_0}(n\epsilon_2)}{4^{n-1}(n-1)!}C_{n-1}^{q}\\
	\times&\exp[ink_0R_0-in\omega_0t+i(nl_0+n-2q-1)\phi].
	\end{aligned}\\
\end{equation}
As a result, at each harmonic order $n$, the diffracted fields for a LP driver contain a sum of $n$ different modes, with topological charges $l_{n,q}=nl_0+n-2q-1$, ($q = 0,1,...,n-1$).
In addition, Eq.~(4) suggests that at harmonic order $n$, the relative amplitude of the $m$th LG modes are proportional to the coefficients $C_{n-1}^{q}$, which explains the azimuthal spectra represented in Fig.~2(h).\\

\begin{figure}[t]
	\centering
	\includegraphics[width=8.5cm]{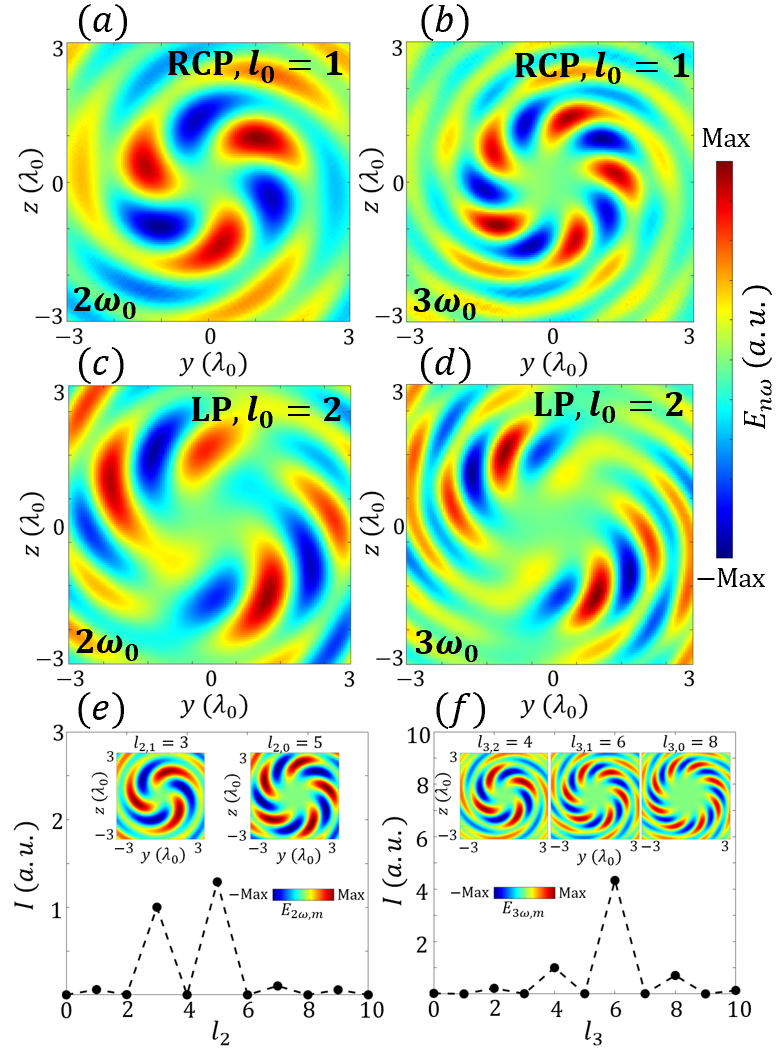}
	\caption{The second and third harmonic fields calculated from our model. (a-b) show the harmonics produced by a RCP LG laser with $l_0=1$ (same as Fig.~1); and (c-d) present the case with a LP LG driver with $l_0=2$ (same as Fig.~2). (e) and (f) are the azimuthal spectra of (c) and (d), respectively. The intensities are normalized by the modes with $l_2=3$ in (e) and $l_3=4$ in (f), the insets depict the filtered LG modes with topological charges specified in the plots.}
	\label{fig:3}
\end{figure}

Although the analytical approach used here are strictly valid only when $a_0\ll1$, the section rules are also applicable to relativistic intensity. 
This is demonstrated in Fig.~3, where we consider harmonic generation in the strongly relativistic regime ($a_0\gg1$). By setting $\delta_{ow} = 0.5\lambda_0$ under the surface-wave-breaking limit \cite{Yi2021}, and solve for $R'(t')$ iteratively, the harmonic fields are calculated numerically from Eq.~(1).

Figures~3(a-b) present the second and third harmonic fields produced by a RCP LG laser with $l_0 = 1$, which agrees very well with Eq.~(3) and the PIC results shown in Fig.~1(d-f). 
Figures~3(c-d) illustrate the second and third harmonic beams generated by a LP LG driver with $l_0 = 2$, which is very similar to Fig.~2(b-c).

Moreover, by performing azimuthal Fourier transformation to Figs.~3(c-d), one obtains the azimuthal mode spectra for the second and third harmonics, presented in Figs.~(e-f). One can see that, as Eq.~(4) predicted, the second harmonics contains two modes $l_{2,0} = 5$ and $l_{2,1} = 3$, and the third harmonic consists of three modes $l_{3,0} = 8$, $l_{3,1} = 6$, and $l_{3,2} = 4$, their relative intensities are in good agreement with PIC results shown Fig.~2(h).\\



In conclusion, we have studied the HHG selection rules for the ROW mechanism driven by a LG pulse. It is found that for a CP driver, single LG mode is produced at each harmonic order, with topological charge $l_n = nl_0+(n-1)\sigma$; when the drive LG pulse is linear polarized, complex spin-orbital momentum interplay occurs. This leads to the generation multiple ($n$) LG modes at $n\omega_0$, with topological charges $l_{n,q}=nl_0+n-2q-1$, where $q = 0,1,...,n-1$.
A theoretical model is derived to explain these observed selection rules based on the ROW model, which agrees very well with the simulations. This work provides fundamental insights into the HHG process via relativistic diffraction. Practically, the harmonic modes with a specific OAM produced by a LP LG driver can be preferentially selected \cite{Yang2017} to produce a frequency comb \cite{Rego2022} or generate UV pulses carrying controlled OAM.\\

\begin{acknowledgments}
	This work is supported by the National Key R$\&$D Program of China (No. 2021YFA1601700), and the National Natural Science Foundation of China (No. 12475246).
\end{acknowledgments}


\begin{thebibliography}{10}
	
\bibitem{Habibovic2024} D. Habibovic, K. R. Hamilton, O. Neufeld, and L. Rego, Nat. Rev. Phys. \textbf{6}, 663 (2024).
\bibitem{Bliokh2015a} {K. Y. Bliokh, F. J. Rodr\'iguez-Frotu\~no, F. Nori, and A. V. Zayats}, Nat. Photon. \textbf{9}, 796 (2015).
\bibitem{Shen2019} {Y. Shen, X. Wang, Z. Xie, C. Min, X. Fu, Q. Liu, M. Gong, and X. Yuan}, Light: Sci. \& Appl. \textbf{8}, 90 (2019).
\bibitem{Berova2013} {N. Berova, P. L. Polavarapu, K. Nakanishi, and R. W. Woody}, {\it Comprehensive Chiroptical Spectroscopy: Instrumentation, Methodologies, and Theoretical Simulations} (Wiley, 2012).
\bibitem{Forbes2018} {K. A. Forbes and D. L. Andrews}, Opt. Lett. \textbf{43}, 435 (2018).
\bibitem{Brullot2016} {W. Brullot, M. K. Vanbel, T. Swusten, and T. Verbiest}, Sci. Adv. \textbf{2}, e1501349 (2016).
\bibitem{Rouxel2022} {J. R. Rouxel, B. R\"osner, D. Karpov, C. Bacellar, G. F. Mancini, F. Zinna, D. Kinschel, O. Cannelli, M. Oppermann, C. Svetina, A. Diaz, J. Lacour, C. David, and M. Chergui}, Nat. Photon. \textbf{16}, 570 (2022).
\bibitem{Mayer2022} {N. Mayer, S. Patchkovskii, F. Morales, M. Ivanov, and O. Smirnova}, Phys. Rev. Lett. \textbf{129}, 243201 (2022).
\bibitem{Ordonez2019} {A. F. Ordonez and O. Smirnova}, Phys. Rev. A \textbf{99}, 043416 (2019).
\bibitem{Wanie2024} {V. Wanie, E. Bloch, E. P. Mansson, L. Colaizzi, S. Ryabchuk, K. Saraswathula, A. F. Ordonez, D. Ayuso, O. Smirnova, A. Trabattoni, V. Blanchet, N. B. Amor, M. C. Heitz, Y. Mairesse, B. Pons, and F. Calegari}, Nature \textbf{630}, 109 (2024).
\bibitem{Gariepy2014} {G. Gariepy, J. Leach, K. T. Kim, T. J. Hammond, E. Frumker, R. W. Boyd, and P. B. Corkum}, Phys. Rev. Lett. \textbf{113}, 153901 (2014).
\bibitem{Geneaux2016} {R. Geneaux, A. Camper, T. Auguste, O. Gobert, J. Caillat, R. Taieb, and T. Ruchon}, Nat. Commun. \textbf{7}, 12583 (2016).
\bibitem{Alon1998} {O. E. Alon, V. Averbukh, and N. Moiseyev}, Phys. Rev. Lett. \textbf{80}, 3743 (1998).
\bibitem{Neufeld2019} {O. Neufeld, D. Podolsky, and O. Cohen}, Nat. Commun. \textbf{10}, 405 (2019).
\bibitem{Tzur2021} {M. E. Tzur, O. Neufeld, A. Fleischer, and O. Cohen}, New J. Phys. \textbf{23}, 103039 (21).
\bibitem{Corkum2007} {P. B. Corkum and F. Krausz}, Nat. Phys. \textbf{3}, 381 (2007).
\bibitem{Bulanov1994} {S. V. Bulanov, N. M. Naumova, and F. Pegoraro}, Phys. Plasmas \textbf{1}, 745 (1994).
\bibitem{Lichters1996} {R. Lichters, J. Meyer-ter-Vehn, and A. Pukhov}, Phys. Plasmas \textbf{3}, 3425 (1996).
\bibitem{Baeva2006} {T. Baeva, S. Gordienko, and A. Pukhov}, Phys. Rev. E \textbf{74}, 046404 (2006).
\bibitem{Dromey2006} {B. Dromey, M. Zepf, A. Gopal, K. Lancaster, M. S. Wei, K. Krushelnick, M. Tatarakis, N. Vakakis, S. Moustaizis, R. Kodama, M. Tampo, C. Stoeckl, R. Clarke, H. Habara, D. Neely, S. Karsch, and P. Norreys}, Nat. Phys. \textbf{2}, 456 (2006).
\bibitem{HG2013} {C. Hernandez-Garcia, A. Picon, J. S. Roman, and L. Plaja}, Phys. Rev. Lett. \textbf{111}, 083602 (2013).
\bibitem{Zhang2015} {X. M. Zhang, B. F. Shen, Y. Shi, X. F. Wang, L. G. Zhang, W. P. Wang, J. C. Xu, L. Yi, and Z. Z. Xu}, Phys. Rev. Lett. \textbf{114}, 173901 (2015).
\bibitem{Yi2021} {L. Yi}, Phys. Rev. Lett. \textbf{126}, 134801 (2021).
\bibitem{Yi2025} {L. Yi}, arXiv:2505.03199 (2025).
\bibitem{Hu2025} {K. Hu, X. Guo, and L. Yi}, arXiv:2505.03215 (2025).
\bibitem{Trines2024} {R. Trines, H. Schmitz, M. King, P. McKenna, and R. Bingham}, Nat. Commun. \textbf{15}, 6878 (2024).
\bibitem{Allen1992} {L. Allen, M. W. Beijersbergen, R. J. C. Spreeuw, and J. P. Woerdman}, Phys. Rev. A \textbf{45}, 8185 (1992).
\bibitem{Shi2014} {Y. Shi, B. F. Shen, L. G. Zhang, X. M. Zhang, W. P. Wang, and Z. Z. Xu}, Phys. Rev. Lett. \textbf{112}, 235001 (2014).
\bibitem{RG2018} {C. Rosales-Guzm\'an, B. Ndagano, and A. Forbes}, J. Opt. \textbf{20}, 123001 (2018).
\bibitem{Bliokh2015PR} {K. Bliokh and F. Nori}, Phys. Rep. \textbf{592}, 1 (2015).
\bibitem{Arber2015} {T. D. Arber, K. Bennett, C. S. Brady, A. Lawrence-Douglas, M. G. Ramsay, N. J. Sircombe, P. Gillies, R. G. Evans, H. Schmitz, A. R. Bell, and C. P. Ridgers}, Plasma Phys. Control. Fusion \textbf{57}, 113001 (2015).
\bibitem{Yang2017} {Y. Yang, G. Thirunavukkarasu, M. Babiker, and J. Yuan}, Phys. Rev. Lett. \textbf{119}, 094802 (2017).
\bibitem{Rego2022} {L. Rego, N. J. Brooks, Q. L. D. Nguyen, J. S. Roman, I. Binnie, L. Plaja, H. C. Kapteyn, M. M. Murnane and C. Hernandez-Garcia}, Sci. Adv. \textbf{8}, 7380 (2022).


\end{thebibliography}


\end{document}